\documentclass[aip,reprint,graphicx]{revtex4-1}
\usepackage{graphicx}

\begin{document}


\title{Efficient Data Averaging for Spin Noise Spectroscopy in Semiconductors} 



\author{Georg M. M{\"u}ller}
\email[Electronic mail: ]{mueller@nano.uni-hannover.de}
\author{Michael R{\"o}mer}

\author{Jens H{\"u}bner}
\author{Michael Oestreich}
\affiliation{Institut f{\"u}r Festk{\"o}rperphysik, Leibniz
Universit{\"a}t Hannover, Appelstra{\ss}e 2, D-30167 Hannover,
Germany}


\date{\today}

\begin{abstract}
Spin noise spectroscopy (SNS) is the perfect tool to investigate
electron spin dynamics in semiconductors  at thermal equilibrium.
We  simulate SNS measurements which utilize real-time fast Fourier transformation instead of an ordinary spectrum analyzer and show that ultrafast digitizers
with low resolution enable surprisingly sensitive, high bandwidth SNS   in the
presence of strong optical background shot noise. The simulations
reveal that optimized input load at the digitizer is  crucial for
efficient spin noise detection while the resolution of the digitizer, i.e., the bit depth, influences the
sensitivity rather weakly.\\
\end{abstract}

\pacs{}

\maketitle 

Spin noise spectroscopy (SNS) has emerged into a powerful tool to
study the spin dynamics in quantum optics and solid-state physics
under equilibrium
conditions.\cite{aleksandrov:jetp:54:64:1981,mueller-2010} The
technique utilizes ever present spin fluctuations to measure the
dynamics of spin ensembles via off-resonant optical Faraday
rotation without disturbing the system. The technique is
 easy to use in atomic gases where the spin noise usually
is of comparable size as optical shot noise.\cite{crooker:nature:431:49:2004}
 The large spin
noise signal in atomic gases results from the long spin relaxation
times and the sharp optical resonances. In semiconductors, typical
spin relaxation times are much shorter and the optical resonances
are significantly broader resulting in a spin noise signal which
is usually orders of magnitude lower than the optical shot noise.
As a consequence, SNS in semiconductors had not been demonstrated
before 2005 \cite{oestreich:prl:95:216603:2005} -- more than 20
years after its first implementation in atom
optics.\cite{aleksandrov:jetp:54:64:1981} Nevertheless, already this first
spin noise measurement on semiconductors indicated the huge
advantage of SNS compared to the traditional optical probes which
inherently disturb the system.

The first SNS experiment on semiconductors was carried out with an
electrical sweeping spectrum analyzer  resulting in a poor
signal-to-noise ratio despite long integration times.
This changed in 2007 by the introduction of real-time fast Fourier
transformation (FFT) spectrum analyzers into SNS which optimized
the data averaging and triggered the current success of
SNS.\cite{romer:rsi:78:103903:2007} Real-time FFT analyzers are by
orders of magnitude more efficient than sweeping spectrum analyzers since
they allow simultaneous detection of spin noise at all frequencies
within the detection bandwidth and thus average over $100\%$ of
the measured data stream.\footnote{A sweeping spectrum analyzer
that was used in Ref.\,\onlinecite{oestreich:prl:95:216603:2005},
e.g., with a bandwidth of 100~MHz and a resolution bandwidth of
0.1~MHz exploits only $\sim 0.1\%$ of the relevant information
carried in the signal.} Nowadays, SNS employing real-time FFT is routinely used to study
the spin dynamics in many bulk and low dimensional
semiconductors.\cite{muller:prl:101:206601:2008,
roemer:apl:94:112105:2009, crooker:prb:79:035208:2009,
romer:prb:81:075216:2010, crooker:prl:104:036601:2010,
mueller:prb:81:121202:2010} However, the technique is  limited to long
spin dephasing times by the bandwidth of the detection system and
in particular by the bandwidth of the electrical analog-to-digital
(A/D) conversion via a digitizer. Up to now,
most SNS experiments have been carried out on systems with spin
dephasing rates below 100~MHz which can be accurately resolved by
fast 16-bit digitizers.\cite{romer:rsi:78:103903:2007,
muller:prl:101:206601:2008, roemer:apl:94:112105:2009,
romer:prb:81:075216:2010, mueller:prb:81:121202:2010} Crooker \textit{et
al.} recently measured spin dephasing rates of a few 100~MHz by means of a
1-GHz digitizer card with a lower effective resolution ($\approx
6$~bit).\cite{crooker:prl:104:036601:2010} A universal application of semiconductor SNS, being
relevant for technological application of SNS
\cite{roemer:apl:94:112105:2009} as well as for fundamental research,
\cite{mueller-2010} certainly demands  an extension of the
available bandwidth exceeding 1~GHz by at least one order of
magnitude. Ultrafast digitizers with a corresponding bandwidth of
up to 13~GHz are commercially available, but show an effective resolution as low as 4~bit at their maximum frequency.

\begin{figure}[tb!]
    \centering
        \includegraphics[width=1.00\columnwidth]{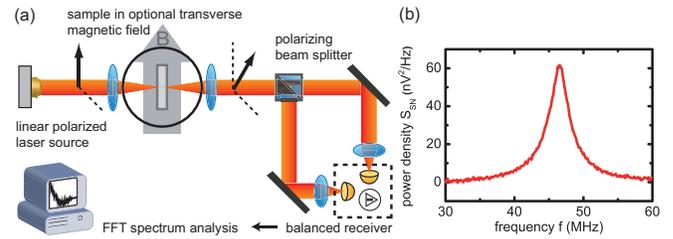}
    \caption{(Color online) (a) Standard setup for semiconductor SNS.
    (b) Typical experimental spin noise power spectrum. The laser shot
    noise background is subtracted.}
    \label{fig:fig4-setup}
\end{figure}
In this letter, we simulate realistic spin noise measurements to
investigate to which extent the low bit depth of fast digitizers
reduces the experimental sensitivity of SNS.
As introduction, we  first elucidate the typical experimental SNS
setup, spin noise spectrum, spin noise power, and background shot
noise level.
Figure~\ref{fig:fig4-setup}~(a) depicts the experimental SNS
setup: Linear polarized probe light is transmitted through the
sample and the acquired stochastic Faraday rotation is measured by
means of a polarizing beam splitter and a balanced photo receiver.
The time signal of the detector is seamlessly digitized, divided
into blocks with fixed size of a power of two, and Fourier
transformed. The resulting Fourier power spectra are added up
for averaging. Figure~\ref{fig:fig4-setup}~(b) depicts a showcase
experimental spin noise spectrum $S_{\mathrm{SN}}(f)
\left[\mathrm{V^2/Hz}\right]$ measured in $n$-doped bulk GaAs at
cryogenic temperatures. The Lorentzian line shape results from a mono-exponential
spin decay and is centered at the electron Larmor frequency
$f_{\mathrm{L}}$ of the precessing spins with a full width at half
maximum proportional to the spin dephasing rate $\Gamma = \pi\,
w_{\mathrm{FWHM}}$. The height of the spin noise peak
$S_{\mathrm{SN}} \left(f_{\mathrm{L}}\right) =2 P_{\mathrm{SN}} \,
\Gamma^{-1} \left(\nu
P_{\mathrm{laser}}\right)^2\,\mathrm{\left[V^2/Hz\right]}$ is
determined by $\Gamma$ and the Faraday rotation noise power
$P_{\mathrm{SN}}\,\mathrm{\left[rad^2\right]}$, which in turn is
given by the number and statistics of the probed spins and the
detuning from the optical resonance.\cite{mueller-2010} The probe laser power is
denoted by $P_{\mathrm{laser}}$ and the conversion gain of the
detector by $\nu\,\mathrm{\left[V/W\right]}$. Optical shot noise
strongly contributes to the total observed noise signal as a white
background noise with a power level $S_{\mathrm{WN}} = 2 \nu^2\,
\hbar\omega_{\mathrm{laser}} P_{\mathrm{laser}} \,
\left[\mathrm{V^2/Hz}\right]$.\footnote{Electrical detector noise further
contributes to the noise background, but only becomes significant  at low probe laser power.}
The low optical density as well as the high spin dephasing rates
in semiconductors result in a very low ratio of peak spin noise
density $S_{\mathrm{SN}}(f_{L})$ to the white background noise level
$S_{\mathrm{WN}}$: $\eta=P_{\mathrm{SN}}\,\Gamma^{-1}\times
P_{\mathrm{laser}}/\hbar\omega_{\mathrm{laser}}$. For typical
semiconductor systems, $\eta$  usually ranges from $10^{-2}$ to
$10^{-4}$ and will be even smaller for prospective room
temperature SNS measurements   in $n$-type bulk GaAs ($\Gamma\gg
1$~GHz)\cite{oertel:apl:93:132112:2007} or for SNS of a single
electron spin  in an optical cavity, where $P_{\mathrm{laser}}$ is
as low as $10\,\mathrm{\mu
W}$ at the balanced
receiver.\cite{berezovsky:science:324:1916:2006}\footnote{See also
Tab.~1 of Ref.~\onlinecite{mueller-2010}.}

\begin{figure}[t!]
    \centering
        \includegraphics[width=1.00\columnwidth]{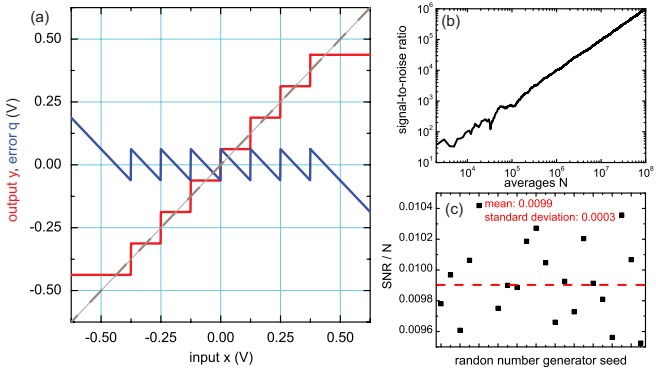}
    \caption{(Color online) (a) Input-output characteristics of a symmetric midrise digitizer ($R=3\,\mathrm{bit}$). (b) Signal-to-noise ratio (SNR) of an exemplary simulated spin noise measurement with $N$ averages. (c) $\mathrm{SNR}/N$ for different random number sets; all other data in this letter is acquired by means of the same set of random numbers.}
    \label{fig:fig2}
\end{figure}
Next, we simulate such SNS measurements for the case of low signal strength
$\eta$ and fast digitizers with low resolution. The key figures of merit of a digitizer are the bit depth $R$ and the sampling rate $f_{\mathrm{S}}$.\cite{JayantNoll1984} The sampling rate limits the bandwidth to $B=f_{\mathrm{S}}/2$ according to the Nyquist-Shannon
theorem.\cite{nyquist:transaiee:47:617:1928,shannon:procire:37:10:1949}
 The bit depth defines the resolution of a digitizer.
Figure~\ref{fig:fig2}(a) depicts the input-output characteristics
of a uniform and symmetric midrise quantizer that is commonly
employed in fast A/D converters. The digital output $y$ of these
digitizers deviates from the analog input signal $x$ by a quantization error
$q=y-x$ which results from the granularity and from the overload
for $|x|>x_{\mathrm{max}}$. We set  $x_{\mathrm{max}}=0.5\,\mathrm{V}$ throughout this letter.  As
long as $q$ can be viewed as distributed independently of $x$, the
granularity results in a standard deviation of $q$ given by
Bennett's formula where it is assumed that $q$ is uniformly distributed and is limited to $-2^{-R}/2<q<2^{-R}/2$, i.e., no overload occurs:\cite{bennett:bellstj:27:446:1948}
\begin{equation}\label{eq:bennett}
\sigma_{q\mathrm{,\,gran}}=2^{-R}/\sqrt{12}\,\mathrm{V}.
\end{equation}

\begin{figure}[t!]
    \centering
        \includegraphics[width=1.00\columnwidth]{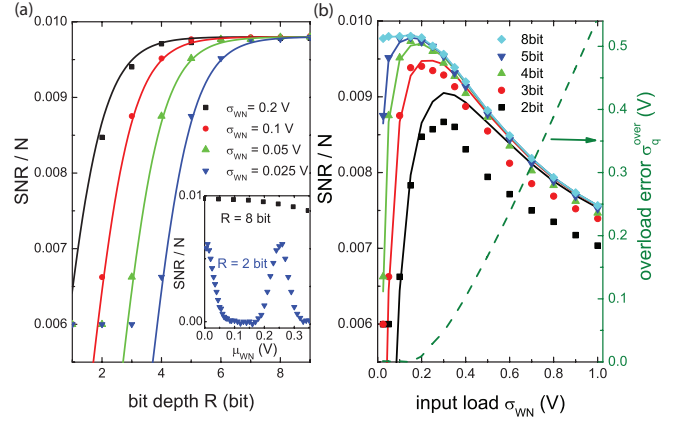}
    \caption{(Color online) (a) $\mathrm{SNR}/N$ ($N=10^8$) as a
    function of bit depth for different voltage loads. The ratio of
    spin noise to shot noise is fixed to $\eta=256\times \alpha^2/\sigma_{\mathrm{WN}}^2=0.01$. The solid lines are calculated by Eqs.~(\ref{eq:bennett}) and (\ref{eq:sigma}) and corrected for the statistic deviation from 0.01. The inset shows $\mathrm{SNR}/N$ as a function of a superimposed DC signal $\mu_{\mathrm{WN}}$ revealing the effective 1-bit quantization at low input load. (b) $\mathrm{SNR}/N$ ($N=10^8$, $\eta=256\times \alpha^2/\sigma_{\mathrm{WN}}^2=0.01$) as a function of  input load for different bit depths. The broken line gives the overload error and is extracted from the difference between the curve in (a) and the values from the simulations for 8~bit.  The solid lines are calculated with these values for $\sigma_{q\mathrm{,\,over}}$ in Eq.
    (\ref{eq:sigma}).}
    \label{fig:fig3}
\end{figure}
In the following simulation of the spin noise measurements, we systematically vary the bit depth of the digitizer and the voltage load at the digitizer input in order to study the influence of the A/D conversion on the experimental sensitivity of SNS. The input consists of $N\times 1024$ samples. The spin noise is represented by a sine waveform with amplitude $\alpha\,\mathrm{[V]}$ and a fixed frequency $f_{\mathrm{L}}=f_{\mathrm{S}}/8$ which is added to random number generated white Gaussian background noise with zero mean and the standard deviation $\sigma_{\mathrm{WN}} \,\mathrm{[V]}$.\footnote{It is carefully assured that the output of the used pseudo random number generator shows no correlations that reduce the efficiency of the averaging process. The periodicity of the employed algorithm exceeds $10^{57}$ samples.} The digital data  is produced by a simulated $R$-bit digitizer corresponding to  Fig.~\ref{fig:fig2}~(a) and blocks of 1024 points are  Fourier transformed via the FFT algorithm.  The resulting power spectra are averaged and yield the spin noise spectrum $S(f) \,\mathrm{\left[V^2/Hz\right]}$. The spin noise signal strength is  measured by the signal-to-noise ratio (SNR) which is extracted from the simulations as $\mathrm{SNR} = \left[S(f_{\mathrm{L}})-\mu_{S} \right]/\sigma_S$, where $\mu_S$ and $\sigma_S$ denote the mean and standard deviation of $S(f)$ with $0 < f \leq B\, ,\, f\ne f_{\mathrm{L}}$, respectively. 
All spin noise power is detected in a single frequency bin $f_{\mathrm{L}}$ and the  magnitude of the spin noise peak is given by $S_{\mathrm{SN}}(f_{\mathrm{L}})=\alpha^2/2\times 512/B$ on top of the shot noise floor of $S_{\mathrm{WN}}= \sigma_{\mathrm{WN}}^2/B$.  All simulation data that is shown in this letter is acquired with fixed relative signal strength  $\eta = S_{\mathrm{SN}}(f_{\mathrm{L}})/S_{\mathrm{WN}} = 256\times \alpha^2/\sigma_{\mathrm{WN}}^2 = 0.01$ since the following conclusions are independent of $\eta$ for $\eta \ll 1$. Keeping $\eta$ fix while varying $\sigma_{\mathrm{WN}}$  corresponds to simultaneous amplification of spin noise and background noise by a voltage amplifier in the experiment. 


Figure~\ref{fig:fig2}~(b) exemplarily shows the signal-to-noise ratio as a function
of the averages $N$ for a certain set of simulation parameters.
The signal-to-noise ratio increases linearly with $N$ and the slope $\mathrm{SNR}/N$
directly gives a measure for the detection sensitivity.
Figure~\ref{fig:fig2}~(c) shows $\mathrm{SNR}/N$ for several
simulations with the same parameter set but with different seeds
for the pseudo random number generator. The theoretically expected
mean value of $0.01$, which is derived below, is well within the error interval
around the mean of $\mu_{\mathrm{SNR}/N}=0.0099(3)$. In
order to cope with the long computing times, we waive statistic
averaging of the simulation results in the following and carry out
all simulation runs with the same set of random numbers to ensure
comparability between different parameters.

Figure~\ref{fig:fig3}~(a) shows $\mathrm{SNR}/N$  as a function of
the bit depth $R$ for  $10^8$ averages and different
$\sigma_{\mathrm{WN}}$. The simulations reveal for an optimal
input load  of
$\sigma_{\mathrm{WN}}=0.1...0.2\,\mathrm{V}$ a significant
decrease of the detection sensitivity only for $R\leq 3\,\mathrm{bit}$.\footnote{As $\eta\ll 1$, the input load is basically given by $\sigma_{\mathrm{WN}}$.}
Lower than optimal input voltages effectively reduce the bit depth
and the granular quantization error becomes significant also for
digitizers with a nominally higher bit depth. Given
Eq.~(\ref{eq:bennett}), the rather weak influence of the bit depth
in Fig.~\ref{fig:fig3}~(a) may at first seem surprising. However,
the white shot noise can be viewed as additive dither to the spin
noise signal which helps to detect spin noise with an amplitude
much smaller than the size of the least significant bit
$2^{-R}\,\mathrm{V}$. The implications of such additive dither on
quantization have already been subject of several investigations in
information theory\cite{schuchmann:ieeetranscom:12:162:1964,
vanderkooy:jaudioengsoc:32:106:1984, gray:ieeetransit:39:805:1993,
wannamaker:ieeetranssp:48:499:2000} and were also considered in connection with data
averaging.\cite{belchamber:talanta:28:547:1981,carbone:ieeetransim:43:389:1994,carbone:ieeeinstrmeas:49:337:2000, koeck:sigproces:81:345:2001, skartlien:ieeetransim:54:103:2005} However, especially the
overload errors, which can be important in SNS, have been greatly
neglected so far. We want to point out that purposely adding
noise in digital data averaging can under certain circumstances
reduce the quantization error, however, the white shot noise in
semiconductor SNS is much higher than the optimal amount of
dither,\cite{skartlien:ieeetransim:54:103:2005} i.e.,
optical shot noise is not a remedy but an obstacle in SNS.

Next, we discuss the granular and the overload error in more
detail. The standard deviation of the spin noise spectrum $S$ is
composed of contributions from the white optical shot noise
$\sigma_{\mathrm{WN}}$, the granular quantization noise
$\sigma_{q\mathrm{,\,gran}}$, and the overload noise
$\sigma_{q\mathrm{,\,over}}$:
\begin{equation}\label{eq:sigma}
\sigma_{S}=\frac{1}{B}\sqrt{
\sigma_{\mathrm{WN}}^2+\sigma_{q\mathrm{,\,gran}}^2+\sigma_{q\mathrm{,\,over}}^2}\,\mathrm{\left[\frac{V^2}{Hz}\right]}.
\end{equation}
The signal-to-noise ratio extracted from the simulations reads accordingly
$\mathrm{SNR}=256\,\alpha^2/(B\sigma_{S}^2) \times N$. First, we point
out that the linear increase of signal-to-noise ratio with $N$ is found in the
simulations for all tested parameter sets. Deviation from this
linear behavior as reported in
Ref.~\onlinecite{skartlien:ieeetransim:54:103:2005} is not
expected due to the larger amount of dither present in our
simulations. We further confirmed the proportionality between the signal-to-noise ratio
and $\alpha^2$ in the simulations over four orders of magnitude from $\eta=10^{-1}$ to $\eta=10^{-4}$.
The granular quantization error at lower bit depths is well
modeled by Eq.~(\ref{eq:bennett}) (solid lines in
Fig.~\ref{fig:fig3}~(a)). However, a significantly smaller
quantization error in the simulations than in Bennett's formula is
found at $R=1\,\mathrm{bit}$ or at effective 1-bit quantization in
the case of low input load. This observation is not surprising
since the assumption for Eq.~(\ref{eq:bennett}) that the quantization error is independent of the input obviously
collapses for 1-bit
quantization.\cite{gray:ieeetransit:36:1220:1990} We note that low
input load at low bit rates results in effective 1-bit
quantization, which theoretically shows a decent detection
sensitivity. However, this theoretical sensitivity is  not of
experimental relevance since even the smallest DC offset
$\mu_{\mathrm{WN}}$ yields a drastic drop of the signal-to-noise ratio as visualized
in the inset of Fig.~\ref{fig:fig3}~(a). On the other hand, input
overload, which becomes significant above
$\sigma_{\mathrm{WN}}=0.2...0.3\,\mathrm{V}$, equally destroys the
efficiency of the spin noise detection even at high bit depth.
Figure~\ref{fig:fig3}~(b) depicts $\mathrm{SNR}/N$ as a function
of $\sigma_{\mathrm{WN}}$ for different $R$. The overload error
for 8~bit (broken line) is extracted from the simulation results
in conjunction with Eqs.~(\ref{eq:bennett}) and (\ref{eq:sigma})
and is in turn utilized for calculating  SNR$/N$ for the different
bit depths (solid lines), revealing that the overload error
slightly depends on $R$. The optimal voltage load, which is found
to be independent of spin noise power, varies from
$\sigma_{\mathrm{WN}}\approx 0.1\,\mathrm{V}$ (8~bit) to
$\sigma_{\mathrm{WN}}\approx 0.3\,\mathrm{V}$ (2~bit).
Interestingly, these values coincide with the literature values
for optimal load in the case of a normally distributed signal in
the absence of
dither.\cite{max:iretransinftheo:6:7:1960,JayantNoll1984}

In conclusion, we simulated spin noise measurements with FFT real-time data acquisition by means of a digitizer with low resolution and determined
the optimal input load at the digitizer for efficient data
averaging. The simulations prove that, at well chosen input load,
fast A/D converters with few effective bits allow SNS
with excellent sensitivity. The simulations pave the way towards
high bandwidth SNS with commercial ultrafast digitizers exceeding
frequencies of 10\,GHz. The experimental
bandwidth that is accordingly accessible with current technology is more than one order of magnitude larger than
the highest currently demonstrated bandwidth.\footnote{For
ultrahigh frequencies, the balanced receiver can be substituted by
a single, ultrafast photodiode, if the classical laser noise is
suppressed, e.g., by a Mach-Zehnder interferometer like setup.}
Such an increase in bandwidth erases one of the main obstacles of
SNS and will allow, e.g., SNS on GaAs at room temperature and, consequently, spatially resolved three-dimensional, non-destructive
doping concentration measurements at room temperature.\cite{roemer:apl:94:112105:2009} We want to
point out, that the results of this letter apply in general to all
similar experiments employing A/D conversion where a small signal with
high bandwidth is detected in the presence of strong background
noise.


The authors thank S.A. Crooker for helpful discussions. This work
was supported by the German Science Foundation (DFG priority
program 1285 `Semiconductor Spintronics'), the Federal Ministry
for Education and Research (BMBF NanoQUIT), and the Centre for
Quantum Engineering and Space-Time Research in Hannover (QUEST).
G.M.M. acknowledges support from the Evangelisches Studienwerk.

\end{document}